\documentclass[]{spie}  

\usepackage{amsmath,amsfonts,amssymb}
\usepackage{graphicx}
\usepackage[colorlinks=true, allcolors=blue]{hyperref}
\usepackage[braket,qm]{qcircuit}

\title{Programming Quantum Computers with Large Language Models}

\author[a]{Elena R. Henderson}
\author[a]{Jessie M. Henderson}
\author[a]{Joshua Ange}
\author[a]{Mitchell A. Thornton}
\affil[a]{Darwin Deason Institute for Cyber Security, Southern Methodist University, \hspace{7em} 6425 Boaz Lane Dallas, TX 75205, USA}

\authorinfo{Send correspondence to JMH: hendersonj@smu.edu. Further author information: ERH: erhenderson@smu.edu, JA: jwange@smu.edu, MAT: mitch@smu.edu}

\begin{document} 
\maketitle

\begin{abstract}
Large language models (LLMs) promise transformative change to fields as diverse as medical diagnosis, legal services, and software development.
One reason for such an impact is LLMs' ability to make highly technical endeavors more accessible to a broader audience.
Accessibility has long been a goal for the growing fields of quantum computing, informatics, and engineering, especially as more quantum systems become publicly available via cloud interfaces.
Between programming quantum computers and using LLMs, the latter seems the more accessible task: while leveraging an LLM's fullest potential requires experience with prompt engineering, any literate person can provide queries and read responses.
By contrast, designing and executing quantum programs---outside of those available online---requires significant background knowledge, from selection of operations for algorithm implementation to configuration choices for particular hardware specifications and providers.
Current research is exploring LLM utility for classical software development, but there has been relatively little investigation into the same for quantum programming.
Consequently, this work is a first look at how well an uncustomized, publicly available LLM can write straightforward quantum circuits.
We examine how well OpenAI's ChatGPT (GPT-4) can write quantum circuits for two hardware providers: the superconducting qubit machines of IBM and the photonic devices of Xanadu.
We find that ChatGPT currently fares substantially better with the former.
\end{abstract}


\section{Introduction}\label{sec:intro}
Large language models (LLMs) promise transformative change to fields as diverse as medical diagnosis, legal services, and software development\cite{WSJ24_1,Katz24,Chen21,Dave23}.
The last five years have seen several publicly-released LLMs, including OpenAI's ChatGPT, Anthropic's Claude, Google's Gemini, Meta's LLaMA, Microsoft's Copilot, and Perplexity's namesake model\cite{GPT23,Anthropic24,Gemini23,Llama23,Copilot23,Perplexity24}.
These tools have shown themselves to be impressive in many contexts\cite{WSJ24_1,Katz24,Chen21,Dave23,Romera24,SB23} and rather inept in others\cite{Schmidt23,NYTimes24}, so their ultimate utility is yet to be determined, including given practical limitations such as finite amounts of training data, computational power, and financing\cite{WSJ24_2}.
In the meantime, the models are noticeably improving, including because they can now pass objective assessments that they previously failed\cite{Bommarito22,Bharade23}.

One reason to expect an eventual wide-ranging impact is LLMs' ability to make highly technical endeavors more accessible to a broader audience.
Accessibility has long been a goal for the growing fields of quantum computing, informatics, and engineering\cite{JC24,AM24,NY11,SV24,MK22}, especially as more quantum systems become publicly available via cloud interfaces.
And because any literate person can provide queries to and read responses from an LLM, there seems room to reduce the amount of theoretical and mechanical background knowledge required to develop programs for various target hardware.
Current research is exploring LLM utility for classical software development, but there has been relatively little investigation into the same for quantum programming\cite{Qiskit23,ZL23}.
Consequently, this work is a first look at how well an uncustomized, publicly available LLM can write straightforward quantum circuits.
We examine how well OpenAI's ChatGPT (GPT-4) can write quantum circuits for two hardware approaches and providers: the superconducting qubit machines of IBM and the photonic devices of Xanadu\cite{IBM,Xanadu}.
Specifically, this paper probes whether ChatGPT can implement quantum programming tasks that have at least one implementation publicly available online, when prompted by a \textit{non-quantum developer}.
If ChatGPT could do so reliably, then it would seem a better quantum programming assistant than search engines (without AI enhancements) have been in the past, providing immediate programming assistance tailored to a user's specific requests.
It is important to clarify that our scope is limited to assessing the efficacy of ChatGPT for individuals new to the quantum space.
While future work ought to look at what is possible with more sophisticated prompts, including those assuming some quantum knowledge, that is not the goal of this work.

The remainder of this paper is as follows: Sec.~\ref{sec:test_programs} introduces the quantum algorithms and computing architectures used to test ChatGPT.
Sec.~\ref{sec:exp_specs} then introduces our process for querying ChatGPT and evaluating its responses, before Sec.~\ref{sec:exp_results} presents the results.
We conclude by situating our findings with avenues for further work.
Before going further, we note that this work assumes substantial background knowledge of discrete quantum computing, continuous quantum computing, and development of large language models.
Because those details---while important---are not required to understand the goals or experimental procedure of this work, we leave detailed descriptions of discrete quantum computing, continuous quantum computing, and the theory of LLMs to other papers\cite{JA22,SB21,SB05,YL24}.

\section{Test Programs}\label{sec:test_programs}
We experiment with IBM's and Xanadu's machines for both theoretical and practical reasons.
First, IBM and Xanadu build quantum computers based upon two different computing paradigms---discrete-variable computing (IBM) and continuous-variable computing (Xanadu)---and so, we test ChatGPT's programming ability for architectures of both types.
Second, we have access to both IBM's and Xanadu's physical hardware and Python packages (Qiskit for IBM\cite{Qiskit} and Strawberry Fields for Xanadu\cite{StrawFields}) that allow for remote execution on physical hardware and for local circuit simulation.
Whether a program is to be run on hardware or simulators determines both circuit logic (\textit{e.g.}, allowable gates and configurations) and implementation syntax (\textit{e.g.}, circuit creation, compilation, and execution calls).

\subsection{Maximally Entangled State Program}\label{sub:prog1}
\begin{figure}[ht]
\begin{center}
\begin{tabular}{c}
\includegraphics[height=1.5cm]{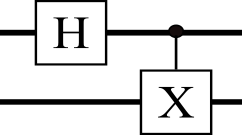}
\end{tabular}
\end{center}
\caption
{\label{fig:max_entangled_qiskit}
A discrete-variable quantum computing circuit that produces a maximally entangled state.  A Hadamard gate acts on the first qubit, and then a Controlled-NOT gate has its control on the first qubit and its target on the second qubit.  This circuit can be programmed using Qiskit.}
\end{figure}

Our first programming request for ChatGPT is both simple and fundamental to quantum computing: implementing a circuit that produces a maximally entangled state.
There are many ways to accomplish this.
Fig. \ref{fig:max_entangled_qiskit} illustrates one straightforward way for the discrete-variable paradigm: formation of a Bell state\cite{IBM_QT}.  Two qubits---both initialized to the state $\ket{0}$---operated on by a Hadamard gate and then a Controlled-Not gate produce a maximally entangled state.

\begin{figure}[ht]
\begin{center}
\begin{tabular}{c}
\includegraphics[height=1.5cm]{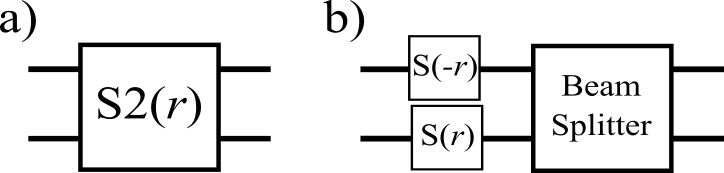}
\end{tabular}
\end{center}
\caption
{\label{fig:max_entangled_SF}
Subfigure a) shows a continuous-variable quantum computing circuit that produces a maximally entangled state if its two-mode squeezer gate applies infinite squeezing amplitude ($r=\infty$).  Subfigure b) shows a decomposition of the two-mode squeezer gate into two local squeezer gates and a 50-50 beamsplitter gate.  Both of these circuits can be programmed using Strawberry Fields.}
\end{figure}

For the continuous-variable case, Fig.~\ref{fig:max_entangled_SF} part a) shows a circuit that produces a maximally entangled state.
A maximally entangled state is produced when a two-mode squeezer gate applies an infinite amount of squeezing (\textit{i.e.}, squeezing amplitude $r$ is infinite) to the two qumodes in the vacuum state\cite{SB05}.
(We note that for the purposes of this paper, complex-valued squeezing amplitude need not be discussed, and hence the real-valued squeezing amplitude $r$.)
This circuit can also be written using the two squeezers and beamsplitter shown in Fig.~\ref{fig:max_entangled_SF} part b)\cite{SB05,Xanadu_QT}.
(All figures in this paper use 50-50 beamsplitters.)

Infinite squeezing amplitude is only possible in theory, and so maximally entangled states are approximated with finite squeezing, where a higher squeezing amplitude means a higher degree of entanglement\cite{SB05}.
Unfortunately, high squeezing amplitudes are difficult to achieve, given phenomena such as photon loss and various sources of noise\cite{HV16}.
The highest measured squeezing amplitude to date is approximately 1.73 (or 15 dB, as measured using quantum noise reduction), similar to the squeezing amplitude of 2 that Xanadu uses in its online documentation for a simulated entangled state\cite{HV16, Xanadu_QT}.
For Xanadu's X8 hardware, the options for squeezing amplitude are only 0 or 1\cite{JA21, Xanadu_X8}.

\subsection{Teleportation Program}\label{sub:prog2}
\begin{figure}[ht]
\begin{center}
\begin{tabular}{c}
\includegraphics[height=1.5cm]{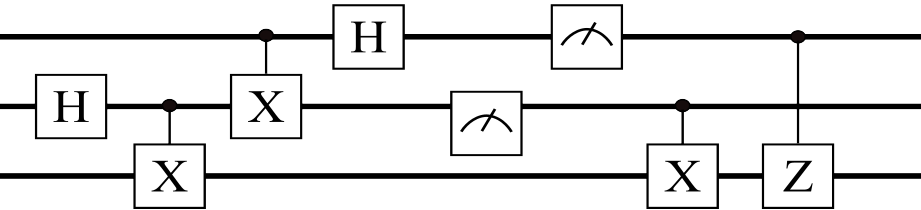}
\end{tabular}
\end{center}
\caption
{\label{fig:teleportation_qiskit}
A discrete-variable computing implementation of the quantum teleportation algorithm.  This circuit can be run using Qiskit.}
\end{figure}

\begin{figure}[ht]
\begin{center}
\begin{tabular}{c}
\includegraphics[height=1.5cm]{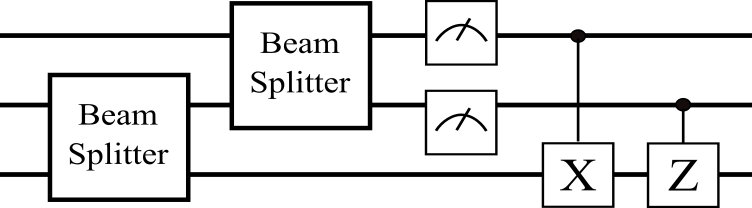}
\end{tabular}
\end{center}
\caption
{\label{fig:teleportation_SF}
A continuous-variable computing implementation of the quantum teleportation algorithm.  This circuit can be run using Strawberry Fields.}
\end{figure}

The second requested program is a well-known quantum algorithm that uses maximal entanglement: teleportation of the state of one qubit/qumode to another.
Implementations of this algorithm for discrete-variable computing and continuous-variable computing are shown in  Figs.~\ref{fig:teleportation_qiskit} and~\ref{fig:teleportation_SF}, respectively\cite{IBM_QT,Xanadu_QT}.

\section{Experiment Specifications}\label{sec:exp_specs}
We consider a total of eight experiments consisting of a requested program (maximal entanglement or quantum teleportation), Python package (Qiskit or Strawberry Fields), and execution method (using a quantum simulator or quantum hardware).
For the maximally entangled state program, we thus have the following four combinations: (1) Qiskit and the AER simulator, (2) Qiskit and the least busy, operational, and available IBM hardware, (3) Strawberry Fields and the Fock backend simulator, and (4) Strawberry Fields and X8 hardware.
We then have the same four options with the teleportation program.
We note the term ``backend'' is synonymous with a specific simulator or hardware instance.
So, for example, the Fock backend simulator is ``a backend," as is a specific hardware machine built by IBM.

For all experiments, we use ChatGPT 4 (without the recently-added memory feature), Strawberry Fields version 0.23.0, and Qiskit version 0.46.1.
The latest version of Qiskit, Qiskit 1.0, is an overhaul of the package that was released just months before this writing, meaning ChatGPT's familiarity is limited to earlier Qiskit versions.

Each of the experiments requires a different query to ChatGPT.
Since ChatGPT's responses vary both with different query formulations and repetitions, we use a standardized query for each experiment, and give ChatGPT multiple opportunities to answer it.
First, query format.
As noted earlier, we do not consider prompt engineering; the goal of this paper is to establish a baseline, and we leave performance-enhancing prompts for future work.
The query for each experiment is given in Sec.~\ref{sec:exp_results} below.

Second, repetitions: for each experiment, we repeat the query five times, each in its own chat session.
Within each chat session, we allow ChatGPT to produce programs with Python implementation errors up to three times.
We define a Python implementation error---hereinafter referred to as an implementation error---as any error that halts program execution.
If a program has an implementation error, we prompt ChatGPT with ``The code produced an error: (last line of the error message)."
ChatGPT tries to explain why this occurred, and usually produces another program that supposedly remedies the issue.
We define a logic error as one that allows the program to run to completion, but does not implement the requested task correctly.
Within each chat session (one ask of a query), we allow ChatGPT one retry if it produces a program with a logic error.
If a program has a logic error, we prompt ChatGPT with ``That program does not produce a maximally entangled state/implement quantum teleportation."
ChatGPT tries to explain why and attempts to produce another program.
If a program has both implementation and logic errors, then ChatGPT is asked to fix each of the former before any of the latter, as implementation errors prevent the program from running to completion.

If implementation errors are resolved (within the allowed maximum of three tries), then ChatGPT is asked (once) to fix any logic errors.
This is the one exception we make in assuming a `non-quantum developer'; to know there is a logic error requires knowing the expected response.
Hence, we give just one attempt for fixing logic errors; we think it interesting to see if ChatGPT's conceptual issues can be resolved with brief informed help, but such was not our primary focus.
If in fixing the logic error, ChatGPT introduces new implementation errors (whether or not there are still logic errors), and if it has not used its three attempts for implementation errors, it is asked to fix those errors.
We apply this iterative structure because ChatGPT changes various aspects of the code when attempting to fix an error, and so can inadvertently fix a different one, such as a logic error when directly asked to fix an implementation error.

In addition to program correctness, we examine ChatGPT's comments within the code.
In all of our experiments, every few lines of code have a one-line comment noting the task executed by those statements (\textit{e.g.}, ``Apply a Hadamard gate to the first qubit") or their purpose (\textit{e.g.}, ``Create entanglement between q[1] and q[2]").
Except where noted in the following sections, these comments correctly describe the goal of the code (whether that code has errors or not).

Finally, we examine program output, namely whether it is informative regarding the circuit task, and whether expected output or output interpretation are explained.
Alongside program output, we assess ChatGPT's descriptions of its programs.
In many cases, these are where ChatGPT provides more detail than the in-code comments, including about important code segments, output interpretation, and quantum concepts.

Each of the tables in the following sections are for one experiment (as indicated by the table caption).
Each row of a table represents one program produced in a certain chat session.
For example, an entry of ``1.3" under the "Query" column means that the program in this row is the third program produced in the first of the five chat sessions.
The ``Implementation Errors" column reports whether the program has implementation errors, and the ``Logic Errors" column reports whether the program has logic errors.
Some tables also have a ``Hardware Errors" column, which reports whether ChatGPT adheres to limitations on the selection and arrangement of circuit operations for a given hardware.
These errors prevent program execution, so are a class of implementation errors.
We report them separately to provide a more granular sense of ChatGPT's current efficacy for quantum programming.

\section{Experimental Results}\label{sec:exp_results}
\subsection{Maximally Entangled State Program Results}\label{sub:prog1_results}
For the first test program, one that produces a maximally entangled state, we ask ChatGPT to ``Write a (Qiskit / Strawberry Fields) program that produces a maximally entangled state and runs using (the AER simulator / the least busy and operational hardware machine / the Fock backend simulator / the X8 hardware)."
The first two options for execution are for Qiskit, and the second two options are for Strawberry Fields.

\begin{table}[ht]
\caption{Logic and implementation errors for the maximally entangled state programs with Qiskit and the AER simulator.}
\label{tab:prog1_qiskit_aer_table}
\begin{center}       
\begin{tabular}{|c|c|c|} 
\hline
\rule[-1ex]{0pt}{3.5ex}  Query & Logic Errors & Implementation Errors  \\
\hline
\rule[-1ex]{0pt}{3.5ex}  1 & No & No \\
\hline
\rule[-1ex]{0pt}{3.5ex}  2 & No & No \\
\hline
\rule[-1ex]{0pt}{3.5ex}  3 & No & No \\
\hline
\rule[-1ex]{0pt}{3.5ex}  4 & No & No \\
\hline
\rule[-1ex]{0pt}{3.5ex}  5 & No & No \\
\hline
\end{tabular}
\end{center}
\end{table} 

\subsubsection{Qiskit with AER Simulator}\label{sub:prog1_qiskit_aer}
When writing programs that use the Qiskit package with the AER simulator, ChatGPT does very well.
As shown in Table~\ref{tab:prog1_qiskit_aer_table}, for each of the five queries, the resulting program contains neither logic nor implementation errors.
In all cases, ChatGPT uses the circuit shown in Fig.~\ref{fig:max_entangled_qiskit} to produce a maximally entangled state.

As for output, for each of the five programs, ChatGPT has circuit measurements printed to the screen, and in all but one case, its description of the code explains how these results would demonstrate the maximal entanglement.
The code descriptions for these programs are accurate and informative, providing a step-by-step guide to the program, including noting the action of both circuit elements (superposition from the Hadamard gate and then entanglement from the CNOT gate).

\begin{table}[ht]
\caption{Logic and implementation errors for the maximally entangled state programs with Qiskit and the least busy hardware.} 
\label{tab:prog1_qiskit_hardware_table}
\begin{center}       
\begin{tabular}{|c|c|c|} 
\hline
\rule[-1ex]{0pt}{3.5ex}  Query & Logic Errors & Implementation Errors  \\
\hline
\rule[-1ex]{0pt}{3.5ex}  1.1 & No & Yes \\
\hline
\rule[-1ex]{0pt}{3.5ex}  1.2 & No & Yes \\
\hline
\rule[-1ex]{0pt}{3.5ex}  1.3 & No & Yes \\
\hline
\rule[-1ex]{0pt}{3.5ex}  2.1 & No & Yes \\
\hline
\rule[-1ex]{0pt}{3.5ex}  3.1 & No & Yes \\
\hline
\rule[-1ex]{0pt}{3.5ex}  4.1 & No & Yes \\
\hline
\rule[-1ex]{0pt}{3.5ex}  4.2 & No & Yes \\
\hline
\rule[-1ex]{0pt}{3.5ex}  4.3 & No & Yes \\
\hline
\rule[-1ex]{0pt}{3.5ex}  5.1 & No & Yes \\
\hline
\rule[-1ex]{0pt}{3.5ex}  5.2 & No & Yes \\
\hline
\rule[-1ex]{0pt}{3.5ex}  5.3 & No & Yes \\
\hline
\end{tabular}
\end{center}
\end{table} 

\subsubsection{Qiskit with Hardware}\label{sub:prog1_qiskit_hardware}
When writing the same program for Qiskit hardware, ChatGPT correctly implements the circuit logic, as shown by the lack of logic errors in Table~\ref{tab:prog1_qiskit_hardware_table}.
But difficulty arises with the hardware implementation: no programs run to completion.
The issue is that all of the programs contain the same obsolete method for loading a hardware backend, and ChatGPT is unable to diagnose this problem, likely given the generic error message that is produced.
(In two programs, ChatGPT made additional errors: neglecting to import a package, and making all hardware machines ineligible due to an unnecessary constraint on hardware parameters added in attempt to fix the recurring error.  But both of these errors were made in the third iteration of a session, and so ChatGPT was not given the opportunity to remedy them.)

ChatGPT's code modifications in the subsequent programs of a chat session are slight, consisting of displaying backend properties for the user and/or raising errors if a backend could not be found.
Indeed, for chat sessions 2 and 3, ChatGPT only provides debugging assistance and no additional programs.
However, it is worth noting two points.
First, except for the two instances described above, when ChatGPT adds code to try and fix the hardware issues, those additions do not introduce new errors.
And second, when ChatGPT does not adjust the code, it does provide an often-lengthy list of possible reasons for the error and suggestions for investigating, including code snippets and advice to look to Qiskit's documentation, support, and forums.

As with the AER simulator programs, ChatGPT's in-code comments and program descriptions for the hardware programs are accurate and informative.
However, except for one instance, ChatGPT does not add comments explaining the output, even though that output would have been measurements qualitatively similar to those from the simulator programs.

\begin{table}[ht]
\caption{Logic and implementation errors for the maximally entangled state programs with Strawberry Fields and the Fock backend.}
\label{tab:prog1_SF_Fock_table}
\begin{center}       
\begin{tabular}{|c|c|c|} 
\hline
\rule[-1ex]{0pt}{3.5ex}  Query & Logic Errors & Implementation Errors  \\
\hline
\rule[-1ex]{0pt}{3.5ex}  1.1 & Yes & Yes \\
\hline
\rule[-1ex]{0pt}{3.5ex}  1.2 & Yes & Yes \\
\hline
\rule[-1ex]{0pt}{3.5ex}  1.3 & Yes & Yes \\
\hline
\rule[-1ex]{0pt}{3.5ex}  2.1 & No & Yes \\
\hline
\rule[-1ex]{0pt}{3.5ex}  2.2 & No & No \\
\hline
\rule[-1ex]{0pt}{3.5ex}  3.1 & Yes & Yes \\
\hline
\rule[-1ex]{0pt}{3.5ex}  3.2 & Yes & No \\
\hline
\rule[-1ex]{0pt}{3.5ex}  3.3 & Yes & Yes \\
\hline
\rule[-1ex]{0pt}{3.5ex}  3.4 & Yes & No \\
\hline
\rule[-1ex]{0pt}{3.5ex}  4.1 & Yes & Yes \\
\hline
\rule[-1ex]{0pt}{3.5ex}  4.2 & Yes & Yes \\
\hline
\rule[-1ex]{0pt}{3.5ex}  4.3 & Yes & No \\
\hline
\rule[-1ex]{0pt}{3.5ex}  4.4 & Yes & No \\
\hline
\rule[-1ex]{0pt}{3.5ex}  5.1 & Yes & Yes \\
\hline
\rule[-1ex]{0pt}{3.5ex}  5.2 & Yes & Yes \\
\hline
\rule[-1ex]{0pt}{3.5ex}  5.3 & Yes & No \\
\hline
\rule[-1ex]{0pt}{3.5ex}  5.4 & Yes & No \\
\hline
\end{tabular}
\end{center}
\end{table} 

\subsubsection{Strawberry Fields with Fock Backend Simulator}\label{sub:prog1_ST_Fock}
When working with Strawberry Fields and the Fock backend simulator, ChatGPT struggles to write a program that produces a maximally entangled state.
As shown in Table~\ref{tab:prog1_SF_Fock_table}, only one of the programs is free from both logic and implementation errors, only two have no logic errors, and only six have no implementation errors.
Almost all of ChatGPT's circuit logic problems stem from the fact that it tries to create the discrete-variable Bell state shown in Fig.~\ref{fig:max_entangled_qiskit}, which is not applicable for the continuous-variable computing of Xanadu.
Indeed, at one point ChatGPT tries to use the discrete gates found in Fig.~\ref{fig:max_entangled_qiskit}, which do not exist for Xanadu programs.
In the few programs where ChatGPT does use squeezing operations, it fails to properly parameterize them or adds other unnecessary gates to the circuit.

Unlike the Qiskit programs, where there are no logic errors, both ChatGPT's circuit logic and implementation details vary widely from program to program in the Xanadu case.
For example, in one program ChatGPT sets a random seed, when no operations in the code require that sort of randomness.
ChatGPT's only justification for this statement is that it allows ``for reproducibility."
However, ChatGPT can self-correct from this mistake; the random seed statement has a syntax error, and once ChatGPT is informed of this, it not only removes the random seed from the program, but also thoroughly explains it is unnecessary.
In other examples, ChatGPT is less successful: it not only experiments with non-existent discrete gates, but misnames Xanadu's phase gate twice in a row, or tries to use object variables from a different simulator than the Fock backend.

As for output, in almost all cases ChatGPT provides the density matrix or the reduced density matrix of the final state.
Information from this matrix can be extracted for interpretation or computation that would illustrate degree of entanglement, but ChatGPT does not explain nor execute any of these options.
In the two instances that ChatGPT explains how to observe maximal entanglement---by examining provided state probabilities---its circuit logic has errors, preventing these descriptions from indicating  success.
ChatGPT also attempts to compute Von Neumann entropy three times, and correctly notes that this metric is a measure of entanglement for a pure bipartite state.
But the execution fails, since ChatGPT either attempts to call a non-existent Strawberry Fields function for computing this metric, or writes its own incorrect computation code that fails to use the storage formatting of density matrices in Strawberry Fields.
Finally, ChatGPT does not describe how Von Neumann entropy relates to degree of entanglement, nor what a user should expect to see.

For nearly all experiments, in-code comments correctly identify what operations are applied (\textit{e.g.}, a beamsplitter gate applied to a qumode).
The only exception is one comment that appears in three programs, where half of the statement applies to a much later line of code.
However, the prose commentary that ChatGPT provides alongside the code is often incorrect.
For example, ChatGPT often claims to be producing Bell states or performing measurements thereof, neither of which is applicable for continuous-variable quantum computing.
Furthermore, even when the commentary is correct, it is vague.
For example, in the circuits that use squeezer gates---and thus are closer to correct---ChatGPT notes that the chosen squeezing value of $1$ is usually sufficient.
But ChatGPT does not explain what this value is sufficient for---approximation of maximal entanglement---or that squeezing amplitude affects degree of entanglement where ``degree" means closeness to maximal entanglement.
Additionally, ChatGPT does not explain that in a simulated program, squeezing amplitude---and thus degree of entanglement---can be increased to values that are not available with physical hardware.

\begin{table}[ht]
\caption{Logic, implementation, and hardware errors for the maximally entangled state programs with Strawberry Fields and X8 hardware.}
\label{tab:prog1_SF_X8_table}
\begin{center}       
\begin{tabular}{|c|c|c|c|} 
\hline
\rule[-1ex]{0pt}{3.5ex}  Query & Logic Errors & Implementation Errors & Hardware Errors  \\
\hline
\rule[-1ex]{0pt}{3.5ex}  1.1 & Yes & Yes & Yes \\
\hline
\rule[-1ex]{0pt}{3.5ex}  1.2 & No & Yes & No \\
\hline
\rule[-1ex]{0pt}{3.5ex}  1.3 & No & No & No \\
\hline
\rule[-1ex]{0pt}{3.5ex}  2.1 & Yes & Yes & Yes \\
\hline
\rule[-1ex]{0pt}{3.5ex}  2.2 & Yes & Yes & Yes \\
\hline
\rule[-1ex]{0pt}{3.5ex}  2.3 & Yes & Yes & Yes \\
\hline
\rule[-1ex]{0pt}{3.5ex}  3.1 & Yes & Yes & Yes \\
\hline
\rule[-1ex]{0pt}{3.5ex}  3.2 & Yes & Yes & Yes \\
\hline
\rule[-1ex]{0pt}{3.5ex}  3.3 & Yes & Yes & Yes \\
\hline
\rule[-1ex]{0pt}{3.5ex}  4.1 & Yes & Yes & Yes \\
\hline
\rule[-1ex]{0pt}{3.5ex}  4.2 & Yes & Yes & Yes \\
\hline
\rule[-1ex]{0pt}{3.5ex}  4.3 & Yes & Yes & Yes \\
\hline
\rule[-1ex]{0pt}{3.5ex}  5.1 & Yes & Yes & Yes \\
\hline
\rule[-1ex]{0pt}{3.5ex}  5.2 & Yes & Yes & Yes \\
\hline
\rule[-1ex]{0pt}{3.5ex}  5.3 & Yes & Yes & Yes \\
\hline
\end{tabular}
\end{center}
\end{table} 

\subsubsection{Strawberry Fields with X8 Hardware}\label{sub:prog1_SF_X8}
Xanadu's X8 hardware---like all contemporary quantum hardware---has limitations on the operations that can be applied and requirements for
the circuit setup\cite{JA21}.
And while Qiskit is now capable of automatic ``transpilation'' such that IBM hardware is susceptible to fewer user-facing limitations, Strawberry Fields is not yet to that point of development.
Consequently, Table~\ref{tab:prog1_SF_X8_table} has the ``Logic Errors" and ``Implementation Errors" from previous sections, as well as ``Hardware Errors," which note whether the circuit ignores hardware requirements.
As noted above, hardware errors are a subset of implementation errors (because they would cause circuit execution to fail), but considering them separately allows for more granularity.
We note that with the gates available on X8 hardware---which include the two-mode squeezer gate in Fig.~\ref{fig:max_entangled_SF} a)---it is possible to produce a maximally entangled state.

As shown in Table~\ref{tab:prog1_SF_X8_table}, in only one instance does ChatGPT produce a program with no logic and no implementation errors.
After being told---via an error message---it cannot use the modes it selected for program 1.1, ChatGPT fixes a missing parameter in program 1.2, and then program 1.3 successfully produces a maximally entangled state.
Additionally, in moving from program 1.1 to program 1.2, ChatGPT fixes an additional logic error without prompting---removing an erroneous beamsplitter gate---an error it makes in nine subsequent programs without acknowledging or fixing.
In all but two cases, ChatGPT's programs have logic errors, the same situation as with the simulator version of the program.

Additionally, ChatGPT fails to acknowledge X8's hardware restrictions in all but two of the programs, contributing to the plethora of implementation errors.
The most common errors are attempting to apply a two-mode squeezer gate to ineligible pairs of modes, and using single-mode squeezer gates and measurements that are not available on X8 hardware\cite{JA21}.

Only one program successfully produces output, and it displays measurement outcomes.
ChatGPT does not provide any explanation of how these measurements could be analyzed to assess whether the circuit produced a maximally entangled state.

ChatGPT's in-code comments are rather pedantic---for example, applying gate \textit{X} to qumode \textit{Y}---and rarely incorrectly state what circuit operations are being applied.
But as with the Fock backend programs, more detailed commentary reflects errors in the circuit, and the same is true for the program descriptions.
For example, in three of the five chat sessions, ChatGPT again states that its programs produce the discrete-variable Bell state in Fig.~\ref{fig:max_entangled_qiskit}, which is as wrong as it is for the simulator programs.
Interestingly, in one chat session, ChatGPT notes that the creation of this Bell state ``isn't straightforward" on X8 hardware---it is in fact not possible---so it will create a ``similar two-mode squeezed state"---which is true in the sense that both are maximally entangled.
In its commentary, ChatGPT sometimes acknowledges that the X8 hardware has restrictions, but then not only ignores them in its programs, but also makes erroneous comments about them, including asserting an ineligible pair of modes could be squeezed, ineligible measurements---that would destroy the entangled state---could be added, squeezer gates would be applied automatically, and a squeezing amplitude of 2 could be used.
Finally, ChatGPT incorrectly describes the purpose of gates it erroneously adds to the circuit.

\subsection{Teleportation Program Results}\label{sec:prog2_results}
To implement quantum teleportation, we ask ChatGPT to ``Write a (Qiskit / Strawberry Fields) program that implements quantum teleportation and runs using (the AER simulator / the least busy and operational hardware machine / the Fock backend simulator / the X8 hardware).
As in Sec.~\ref{sub:prog1}, the first two options for execution are for Qiskit, and the second two options are for Strawberry Fields.

To test the correctness of a circuit implementation of quantum teleportation, one can add a single gate to the circuit (and examine the proper output)\cite{IBM_QT}.
Considering the three-qubit/qumode circuits in Figs.~\ref{fig:teleportation_qiskit} and~\ref{fig:teleportation_SF}, one simply adds a gate to the third qubit/qumode that is the inverse of the gate that was used to produce the state to be teleported on the first qubit/qumode.
After circuit execution, if the state of the first qubit/qumode was successfully teleported to the the third qubit/qumode, the latter should be measured in the state $\ket{0}$ (in the computational basis for a Qiskit-specified circuit, and in the Fock, or number, basis for a Strawberry Fields-specified circuit).
We use this inversion test---in addition to comparing circuit logic to known correct teleportation circuits---to evaluate ChatGPT's teleportation programs.

Because of sample noise, one must run an `inverse test' circuit multiple times.
In our experiments, we run each `inverse test' circuit 1,024 times.
Qiskit allows the user to select multiple circuit executions (called ``shots''), while Strawberry Fields does not, so we manually loop over those programs.
Since IBM's own teleportation example circuit---when executed 1,024 times---does not produce any measurement results where the third qubit is not in state $\ket{0}$, we require the same of ChatGPT's circuits in order to pass the inverse test for teleportation.

For Strawberry Fields programs, we examine the number state probabilities of the third qumode.
Xanadu's own teleportation example circuit---when executed 1,024 times---has on average $85.91\%$ probability that the state of the third qumode will be measured in number state $\ket{0}$.
And so, in assessing whether ChatGPT's Strawberry Fields circuits pass the inverse test for teleportation, we require that the third qumode have at least an $85.91\%$ probability that it will be measured in state $\ket{0}$.

Finally, the inverse test is a necessary, not sufficient, criteria for successful teleportation, and so we always examine circuit logic, as there are five Strawberry Fields programs where ChatGPT makes logic errors, but the inverse test passes (\textit{e.g.,} where measurement values are incorrectly set to zero).

\begin{table}[ht]
\caption{Logic and implementation errors for the teleportation programs with Qiskit and the AER simulator.} 
\label{tab:prog2_qiskit_aer_table}
\begin{center}       
\begin{tabular}{|c|c|c|} 
\hline
\rule[-1ex]{0pt}{3.5ex}  Query & Logic Errors & Implementation Errors  \\
\hline
\rule[-1ex]{0pt}{3.5ex}  1.1 & Yes & No \\
\hline
\rule[-1ex]{0pt}{3.5ex}  1.2 & No & No \\
\hline
\rule[-1ex]{0pt}{3.5ex}  2 & No & No \\
\hline
\rule[-1ex]{0pt}{3.5ex}  3.1 & Yes & No \\
\hline
\rule[-1ex]{0pt}{3.5ex}  3.2 & Yes & No \\
\hline
\rule[-1ex]{0pt}{3.5ex}  4 & No & No \\
\hline
\rule[-1ex]{0pt}{3.5ex}  5.1 & Yes & No \\
\hline
\rule[-1ex]{0pt}{3.5ex}  5.2 & No & No \\
\hline
\end{tabular}
\end{center}
\end{table} 

\subsubsection{Qiskit with AER Simulator}\label{sub:prog2_qiskit_aer}
For programs using Qiskit and the AER simulator, ChatGPT performs fairly well, and is able to recover from all but one of its mistakes (in the allowed attempts---recall that ChatGPT is allowed one try to correct a logic error and three tries to correct an implementation error).
The circuit logic mistakes all arise from ChatGPT's improperly applying the final measurement operations at the end of the circuit.

As for program output, in all of the programs but one, measurement counts are displayed.
(In one program, the output is a state vector, which, as the combined final state of all three qubits written as one vector, is not particularly useful for assessing whether teleportation is successful.)
In two cases this output is not informative, as the third qubit---which should contain the teleported state---is not measured, and so could not be examined.

In all but two of the programs, ChatGPT provides a step-by-step overview of the circuit logic with correct in-code comments.
But ChatGPT's explanation of the program output varies in quality.
In about half of the cases, it is unhelpfully vague, stating that the output would show teleportation, but not explaining why.
In the other half, it describes how the measurement counts should demonstrate the teleportation.

\begin{table}[ht]
\caption{Logic and implementation errors for the teleportation programs with Qiskit and the least busy hardware.} 
\label{tab:prog2_qiskit_hardware_table}
\begin{center}       
\begin{tabular}{|c|c|c|} 
\hline
\rule[-1ex]{0pt}{3.5ex}  Query & Logic Errors & Implementation Errors  \\
\hline
\rule[-1ex]{0pt}{3.5ex}  1.1 & No & Yes \\
\hline
\rule[-1ex]{0pt}{3.5ex}  2.1 & No & Yes \\
\hline
\rule[-1ex]{0pt}{3.5ex}  2.2 & No & Yes \\
\hline
\rule[-1ex]{0pt}{3.5ex}  3.1 & Yes & Yes \\
\hline
\rule[-1ex]{0pt}{3.5ex}  3.2 & Yes & Yes \\
\hline
\rule[-1ex]{0pt}{3.5ex}  3.3 & N/A & Yes \\
\hline
\rule[-1ex]{0pt}{3.5ex}  4.1 & No & Yes \\
\hline
\rule[-1ex]{0pt}{3.5ex}  4.2 & Yes & Yes \\
\hline
\rule[-1ex]{0pt}{3.5ex}  4.3 & No & Yes \\
\hline
\rule[-1ex]{0pt}{3.5ex}  5.1 & No & Yes \\
\hline
\rule[-1ex]{0pt}{3.5ex}  5.2 & No & Yes \\
\hline
\rule[-1ex]{0pt}{3.5ex}  5.3 & No & Yes \\
\hline
\end{tabular}
\end{center}
\end{table} 

\subsubsection{Qiskit with Hardware}\label{sub:prog2_qiskit_hardware}
ChatGPT's performance with programs for Qiskit running on quantum hardware is both similar and different to its performance with the AER simulator programs.
Given the number of programs, ChatGPT does not make many logic errors, and is able to recover from one of them.
(It resolves the logic error in program 4.2 while attempting to resolve a persistent implementation error.)
As with the AER simulator version of the program, the logic errors occur in the measurement-based parametrization of the final control gates.
We note that program 3.3 in Table~\ref{tab:prog2_qiskit_hardware_table} has logic errors listed as not applicable.
ChatGPT wrote this program so the user could determine whether the circuit was the source of a persistent implementation error discussed below; the code (correctly) implements only a maximally entangled state, but it does not resolve the error.

Although the programs are thus reasonably correct from the logic error perspective, in contrast to the simulator programs, every one of the programs in Table~\ref{tab:prog2_qiskit_hardware_table} has implementation errors, meaning none runs to completion.
These errors were the same as those for the maximally entangled state hardware programs of Section \ref{sub:prog1_qiskit_hardware} and have the same cause: an outdated method for connecting to IBM hardware.
As in the maximally entangled state programs, ChatGPT is unable to resolve this error, likely given the non-specificity of the resulting error message.

Also similar to the the maximally entangled state Qiskit programs running on hardware, ChatGPT suggests how to debug and error-check the persistent error.
Except for two cases, this commentary is much less extensive and detailed than for the case of Sec.~\ref{sub:prog1_qiskit_hardware}.
The two programs for which ChatGPT does provide comparable advice, including example code snippets, are those for which ChatGPT does not provide an iterated version of the code.
Even when ChatGPT does make adjustments to code, they are rather light, again following the trends of the maximally entangled state hardware programs.
(In two instances, ChatGPT modifies the circuit logic and introduces new implementation errors.)
Indeed, on multiple occasions, ChatGPT claims to have implemented additional error checking and improvements, but makes only small, often irrelevant modifications, such as printing the name of the target hardware.

When the programs in Table~\ref{tab:prog2_qiskit_hardware_table} run to completion, they often produce useless output, as the circuits do not measure the qubit containing the teleported state, meaning it is not included in displayed results.

Finally, we note that ChatGPT's commentary and program descriptions are informative and detailed, as is often the case for the Qiskit programs.

\begin{table}[ht]
\caption{Logic and implementation errors for the teleportation programs with Strawberry Fields and the Fock backend.}
\label{tab:prog2_SF_Fock_table}
\begin{center}       
\begin{tabular}{|c|c|c|} 
\hline
\rule[-1ex]{0pt}{3.5ex}  Query & Logic Errors & Implementation Errors \\
\hline
\rule[-1ex]{0pt}{3.5ex}  1.1 & Yes & Yes \\
\hline
\rule[-1ex]{0pt}{3.5ex}  1.2 & Yes & No \\
\hline
\rule[-1ex]{0pt}{3.5ex}  1.3 & Yes & Yes \\
\hline
\rule[-1ex]{0pt}{3.5ex}  1.4 & Yes & No \\
\hline
\rule[-1ex]{0pt}{3.5ex}  2.1 & Yes & Yes \\
\hline
\rule[-1ex]{0pt}{3.5ex}  2.2 & Yes & Yes \\
\hline
\rule[-1ex]{0pt}{3.5ex}  2.3 & Yes & Yes \\
\hline
\rule[-1ex]{0pt}{3.5ex}  3.1 & Yes & Yes \\
\hline
\rule[-1ex]{0pt}{3.5ex}  3.2 & Yes & Yes \\
\hline
\rule[-1ex]{0pt}{3.5ex}  3.3 & Yes & No \\
\hline
\rule[-1ex]{0pt}{3.5ex}  3.4 & Yes & Yes \\
\hline
\rule[-1ex]{0pt}{3.5ex}  4.1 & Yes & Yes \\
\hline
\rule[-1ex]{0pt}{3.5ex}  4.2 & Yes & No \\
\hline
\rule[-1ex]{0pt}{3.5ex}  4.3 & Yes & No \\
\hline
\rule[-1ex]{0pt}{3.5ex}  5.1 & Yes & Yes \\
\hline
\rule[-1ex]{0pt}{3.5ex}  5.2 & Yes & No \\
\hline
\rule[-1ex]{0pt}{3.5ex}  5.3 & Yes & No \\
\hline
\end{tabular}
\end{center}
\end{table} 

\subsubsection{Strawberry Fields with Fock Backend Simulator}\label{sub:prog2_SF_Fock}
For teleportation programs written with Strawberry Fields and running with the Fock backend simulator, ChatGPT is unable to produce an error-free program.
All of the programs have logic errors (and many also have implementation errors), as shown in Table~\ref{tab:prog2_SF_Fock_table}.
Not only do ChatGPT's adjustments fail to eliminate targeted errors, but they also introduce new errors, such as in programs 1.3 and 3.4.
This is often the consequence of ChatGPT adjusting code that has no bearing on the error, and usually, this adjustment is without explanation.

Additionally, the errors ChatGPT makes are significantly more wide-ranging than those it makes for the Qiskit teleportation programs, which are almost always in the same part of the circuit logic or setup code.
The errors in the teleportation/Strawberry Fields/Fock backend simulator case include incorrect parametrization of at least one of the circuit operations---squeezers, beamsplitters, measurements, and/or conditional gates (as shown in Fig.~\ref{fig:teleportation_SF})---in every circuit, and also incorrect gate ordering or selection altogether.
Additionally, in the fourth chat session, ChatGPT twice attempts to use the incorrect maximally entangled state logic discussed in Sec.~\ref{sub:prog1_ST_Fock}, since forming a maximally entangled state is the first step of the teleportation circuit.
Furthermore, ChatGPT hallucinates object attributes or sets measurements arbitrarily to zero---four times not telling the user why, and twice telling the user that these measurements might or did need to be adjusted, even though it could have made those adjustments itself.
Finally, ChatGPT also produces a few errors that might be fixable by even a novice programmer: in each chat, ChatGPT either incorrectly provides complex parameters to an operation (the ``Coherent" operation used to set a qumode state), and/or neglects to import the NumPy library.

As discussed earlier in this section, demonstrating successful teleportation is most straightforward with an additional copy of the circuit.
For all programs, ChatGPT chooses the reduced density matrix of the third qumode's state as output.
(The state of this qumode should be the same as that of the first qumode at the beginning of the circuit.)
Also as discussed above, density matrices hold important information, but ChatGPT again provides no explanation of how to interpret them or how to use this output to assess successful teleportation.

As with many other Strawberry Fields programs, additional details from in-code comments are often inaccurate.
For example, gate parameters are incorrectly described (selecting which basis a measurement is in), and measurements are said to come from qumode states but are actually set to zero.
Furthermore, ChatGPT repeats the mistake of the maximal entanglement programs with Xanadu, referring to Bell states and measurements thereof when those exist only for discrete quantum computing.

ChatGPT's additional description is a step-by-step discussion of the program that in some cases has a few more details than the in-code comments.
But one prominent omission is the lack of explanation regarding the choice of squeezing amplitude.
When a program uses a two-mode squeezer gate, ChatGPT properly selects a squeezing amplitude of $1$.
But in the majority of programs, where ChatGPT uses two single-mode squeezer gates---which are not available on X8 hardware---it selects squeezing amplitudes of $-2$ for both.
First of all, one of the squeezers should have a positive amplitude and the other a negative amplitude, as shown in Fig.~\ref{fig:max_entangled_SF} part b).
Second, while $2$ is the squeezing amplitude for Xanadu's own teleportation example---which runs with a simulator---this value is not available for X8 hardware\cite{Xanadu_QT}.
As with the maximally entangled state programs, ChatGPT does not comment on how squeezing amplitude relates to maximal entanglement.
In the single case that ChatGPT mentions squeezing amplitude, it states that, ``physical constraints (like squeezing levels and detection efficiencies) are adequate for the demonstration," when ``detection efficiencies" are neither a parameter in the program nor something the user can control.

\begin{table}[ht]
\caption{Logic, implementation, and hardware errors for the teleportation programs with Strawberry Fields and X8 hardware.} 
\label{tab:prog2_SF_X8_table}
\begin{center}       
\begin{tabular}{|c|c|c|c|} 
\hline
\rule[-1ex]{0pt}{3.5ex} Query & Logic Errors & Implementation Errors & Hardware Errors  \\
\hline
\rule[-1ex]{0pt}{3.5ex}  1.1 & Yes & Yes & Yes \\
\hline
\rule[-1ex]{0pt}{3.5ex}  1.2 & Yes & Yes & Yes \\
\hline
\rule[-1ex]{0pt}{3.5ex}  1.3 & Yes & Yes & Yes \\
\hline
\rule[-1ex]{0pt}{3.5ex}  2.1 & Yes & Yes & Yes \\
\hline
\rule[-1ex]{0pt}{3.5ex}  2.2 & Yes & Yes & Yes \\
\hline
\rule[-1ex]{0pt}{3.5ex}  2.3 & Yes & Yes & Yes \\
\hline
\rule[-1ex]{0pt}{3.5ex}  3.1 & Yes & Yes & Yes \\
\hline
\rule[-1ex]{0pt}{3.5ex}  3.2 & Yes & Yes & Yes \\
\hline
\rule[-1ex]{0pt}{3.5ex}  3.3 & Yes & Yes & Yes \\
\hline
\rule[-1ex]{0pt}{3.5ex}  4.1 & Yes & Yes & Yes \\
\hline
\rule[-1ex]{0pt}{3.5ex}  4.2 & Yes & Yes & Yes \\
\hline
\rule[-1ex]{0pt}{3.5ex}  4.3 & Yes & Yes & Yes \\
\hline
\rule[-1ex]{0pt}{3.5ex}  5.1 & Yes & Yes & Yes \\
\hline
\rule[-1ex]{0pt}{3.5ex}  5.2 & Yes & Yes & Yes \\
\hline
\rule[-1ex]{0pt}{3.5ex}  5.3 & Yes & Yes & Yes \\
\hline
\end{tabular}
\end{center}
\end{table} 

\subsubsection{Strawberry Fields with X8 Hardware}\label{sub:prog2_SF_X8}
Asking ChatGPT to write a quantum teleportation program for running on X8 hardware is a bit of a trick question, in that the operations currently available on X8 hardware are not sufficient to program the algorithm.
Yet ChatGPT does not acknowledge this, but rather writes programs that uniformly violate hardware restrictions.
ChatGPT almost always warns that its program would or may need adjustment given hardware constraints, but it does not specify the constraints themselves.
As an example, one of the most common errors is applying measurements that are not available for X8 hardware.
Furthermore, ChatGPT makes logic errors by arbitrarily setting the measured values or leaving out the gates that depend upon these measurements entirely.
While ChatGPT often notes that it is making assumptions about measurements, it does not specify what those assumptions are, and fails to note that the measurements applied are not even available.
In fact, in two programs, it asserts the contrary to be true by failing to recognize three operations are in fact the same type of measurement, two of them being specifically parameterized versions of the third.

Additionally, ChatGPT's descriptions of the measurement issues are often confusing; it characterizes its artificial setting of measurements as ``simplifying" programs, instead of noting that this makes programs logically incorrect.
Finally, the logic errors of every program are often the same as those of the simulator teleportation programs, with additional, more blatant problems, such as attempted corrections to a quantum state after the completion of circuit execution (something that also causes an implementation error).

The output ChatGPT tries to produce varies widely, with some programs displaying no output at all.
Much of the output could only be produced using programs run with simulators, and even then, there is no explanation of expected output or interpretation thereof.

The in-code comments are very similar to those in the teleportation programs for simulation, and so are the additional program descriptions, including ChatGPT's erroneous inclusion of Bell states and measurements.
ChatGPT's program descriptions contain spurious advice such as the assertion that users should---and can, within the Strawberry Fields framework---check for and adjust things like photon loss (from measurement detectors), device imperfections, and calibration issues.
In fact, the user can control none of these parameters.
In another instance, ChatGPT states, ``more error correction" would need to be added to the program, which at the time contained none, at least from the user's programmatic perspective.
And as described above, ChatGPT almost always acknowledges its programs may need adjustments for hardware, but provides no details.

\section{Discussion}\label{sec:discussion}
Section~\ref{sec:exp_results} is rather information dense, so in this section, we briefly summarize our findings with regard to our test metrics for ChatGPT: logic errors, implementation errors (including hardware errors), program output, and program commentary and discussion.

Beginning with circuit logic, ChatGPT performs quite well with Qiskit for our two test programs.
For the simple maximally entangled state, ChatGPT makes no logic errors in any program, and with the more difficult teleportation circuit, logic errors are relatively rare, and ChatGPT can correct almost all of them.
By contrast, ChatGPT's Strawberry Field programs are rife with logic errors, even for the simple maximally entangled state program, and ChatGPT is rarely able to remedy them.
As for implementation errors---ignoring hardware for a moment, which presents a struggle for ChatGPT regardless of programming package---Qiskit simulator programs have none, while Strawberry Fields programs have several.
And so, ChatGPT has lesser prowess with both Xanadu circuit logic and Strawberry Fields implementation than with IBM hardware and Qiskit implementation.

For both Qiskit and Xanadu programs, ChatGPT struggles with hardware setup code, which is not surprising.
Qiskit recently changed its hardware-interface syntax, and historically has changed this frequently in different versions of the package.
As Qiskit becomes a more stable package, we will likely see fewer syntax changes, and therefore improved ChatGPT efficacy.
As for Xanadu, ChatGPT ignores hardware restrictions in all but two cases, likely because of a lack of hardware programming examples online.

Given that ChatGPT performs better with Qiskit circuit logic and implementation, it is no surprise that its commentary on both---for comments in the code and additional program descriptions---is more accurate and detailed.
One notable failing with both Qiskit and Strawberry Fields programs is a lack of commentary on programming output, and, especially for Strawberry Fields programs, a lack of useful output to begin with.

\section{Conclusions and Future Work}\label{sec:conclusion}
This paper is a first look at how well one publicly available LLM can program quantum circuits.
Search engines (without added AI) have long been a source of information for the novice quantum programmer, and so a natural first question is how well ChatGPT's quantum programming assistance compares to that of readily available internet sources presenting straightforward sample circuits.
Such sources include the documentation and sample code from both IBM/Qiskit and Xanadu/Strawberry Fields, but as Xanadu's efforts are newer than IBM's, there is significantly more material available on programming with the latter, including years of contributions from non-official sources, such as Stack Exchange and Medium.

This relative difference in available information is consistent with our results, which suggest that ChatGPT is---all things considered---better equipped to provide quantum circuits using Qiskit for IBM backends than circuits using Strawberry Fields for Xanadu backends.
While ChatGPT can replicate the maximally entangled state and teleportation Qiskit circuits found repeatedly on the internet, it could not consistently replicate either test circuit using Strawberry Fields.
So, searching online---at this point largely for Xanadu's own documentation---is clearly a superior route for new quantum programmers than querying ChatGPT.

Of course this work is just the beginning of an investigation of LLMs and quantum programming.
Because Strawberry Fields currently seems the more challenging target, the next step for assessing ChatGPT's abilities would be to experiment with Qiskit circuits for additional algorithms, including those not as readily available online.
Additional future work includes testing with LLMs mentioned in Sec.~\ref{sec:intro}, considering other quantum computing hardware and software providers\cite{QInsider23_1, QInsider22_3}, and better tailoring queries through prompt engineering.

\acknowledgments      
 
We gratefully acknowledge the use of both Xanadu's Quantum Cloud and IBM Quantum services for this work.
All views are those of the authors, and do not reflect the official policy or position of Xanadu, IBM, or their respective teams.

\bibliography{references} 
\bibliographystyle{spiebib} 

\end{document}